\documentclass[12pt]{article}
\usepackage{putex}
\usepackage{graphicx}
\usepackage{latexsym,amsmath,amsfonts,amssymb,cite}
\usepackage{bbm}
\numberwithin{equation}{section}

\def\btab{\begin{table}[h] \begin{center} \begin{tabular}{l lp{3in}}}
      \def\etab{\end{tabular} \end{center} \end{table}}
\def\btabm{\begin{center} \begin{tabular}}
    \def\etabm{\end{tabular} \end{center}}

\def\D{{\Delta}}

\def\CF{{\cal F}}

\def\CO{{\cal O}}

\begin{document}

\title{Relevant Perturbation of Entanglement Entropy and Stationarity}

\authors{Tatsuma Nishioka}

\institution{IAS}{
School of Natural Sciences, Institute For Advanced Study, Princeton NJ 08540, USA}

\abstract{
A relevant perturbation of the entanglement entropy of a sphere is examined holographically near the UV fixed point.
Varying the conformal dimension of the relevant operator, we obtain three different sectors: 1) the entanglement entropy is stationary and the perturbative expansion is well-defined with respect to the relevant coupling, 2) the entropy is stationary, but the perturbation fails, 3) the entropy is neither stationary nor perturbative.
We compare our holographic results with the numerical calculation for a free massive scalar field in three-dimensions, and find a qualitative agreement between them.
We speculate that these statements hold for any relevant perturbation in any  quantum field theory invariant under the Poincar{\'e} symmetry.
}

\date{May 2014}
\maketitle

\tableofcontents
%

\section{Introduction}
Entanglement entropy plays a role of a measure of degrees of freedom in quantum field theories.
In two-dimensions, an alternative proof of the Zamolodchikov's $c$-theorem \cite{Zamolodchikov:1986gt} is provided through
the entropic $c$-function, defined by $c_E(R)=3R S'(R)$ with the entanglement entropy $S(R)$ of an interval of width $R$, which is a monotonically decreasing function ($c_E'(R)\le 0$) under any renormalization group (RG) flow \cite{Casini:2004bw}.
A similar statement in three-dimensions, known as the $F$-theorem \cite{Jafferis:2011zi,Klebanov:2011gs}, is also substantiated by the monotonicity ($\CF'(R)\le 0$) \cite{Casini:2012ei} of the renormalized entanglement entropy (REE) \cite{Liu:2012eea} 
\begin{align}\label{REE}
\CF(R) = R S'(R) - S(R) \ ,
\end{align}
where $S(R)$ is the entanglement entropy of a disk of radius $R$.
The strong subadditivity \cite{Lieb:1973cp} and Lorentz invariance are crucial to prove the inequalities in both cases.
Similar proposals are presented for higher dimensions by replacing the disk with codimension-two spheres in \cite{Myers:2010tj,Klebanov:2011gs,Liu:2012eea} (see also \cite{Jafferis:2012iv,Fei:2014yja} for the five-dimensional examples).

The importance of the Zamolodchikov's $c$-function stems from its stationarity at RG fixed points
against a relevant perturbation 
\begin{align}
I_\text{CFT} ~ \rightarrow I_\text{CFT} + \lambda \int d^d x \sqrt{g} \,\CO (x) \ ,
\end{align}
where $\lambda$ is the coupling constant of the operator $\CO$ with conformal dimension $\D$ less than $d$.
The derivative of the $c$-function with respect to the relevant coupling $\lambda$ is always proportional to the beta-function that always vanishes at the fixed point \cite{Zamolodchikov:1986gt}.
The question has been raised by \cite{Klebanov:2012va} as to whether the entropic $c$-function and REE have the same property, namely they are stationary at the fixed points or not. 
For a free massive scalar field, the exact and numerical results are obtained for $c_E$ \cite{Casini:2005zv} and $\CF$ \cite{Klebanov:2012va}, respectively, and 
both turn out to be non-stationary at the UV fixed points.
It has not been elucidated if it is peculiar to a free massive scalar theory so far 
due to the lack of the exact results of the entanglement entropy for general interacting field theories (refer to e.g. \cite{Casini:2009sr} for free quantum field theories).

Recently, general aspects of the perturbation of entanglement entropy has been explored in \cite{Rosenhaus:2014woa} where the leading correction of the relevant deformation is shown to be the second order, $\delta S = O(\lambda^2)$.\footnote{In \cite{Rosenhaus:2014woa}, the perturbation of the reduced density matrix $\delta\rho$ is assumed to commute with the unperturbed one $\rho$, $[\rho, \delta\rho]=0$, which does not hold in general. We thank Y.\,Nakaguchi for sharing this issue with us.}
The stationarity automatically follows at the UV fixed point ($\lambda=0$) as long as the perturbative expansion is well-defined, namely the entropy can be written as a power series of $\lambda$ with finite coefficients.
The scalar field examples, however, contradict with this general argument and there remains a puzzle yet.

In this letter, we tackle the problem of the stationarity from the holographic viewpoint. \footnote{See \cite{Kim:2014yca} for a related work where the stationarity of the REE in the mass-deformed ABJM theory is studied holographically using Lin-Lunin-Maldacena geometries \cite{Lin:2004nb}.}
To be concrete, we study the entanglement entropy $S(R)$ of a sphere of radius $R$ in $d$-dimensions perturbed by a relevant operator.
The spherical entangling surface is embedded in a flat space
\begin{align}\label{flat}
ds^2 = dt^2 + dr^2 + r^2 d\Omega_{d-2}^2 \ ,
\end{align}
at $t=0$ and $r=R$.
We shall be interested in how it behaves near the UV fixed point as we vary the conformal dimension $\D$ of the operator.
We will see that the perturbation should fail for an operator with $\D\le d/2$ and the non-stationarity further appears for $\D \le d/3$.
This resolves the aforementioned paradox between the general argument and the scalar examples.

The organization of this letter is as follows. In section \ref{ss:NumericEE}, we revisit the example of a free massive scalar field in three-dimensions and present the numerical result whose technical details are available in appendix \ref{ss:app}.
We confirm that the (renormalized) entanglement entropy is linear to the mass squared, i.e. not stationary at the UV fixed point.
This is a nontrivial and calculable example that indicates both non-stationarity and a failure of the perturbation.

Section \ref{ss:Holographic} deals with the holographic entanglement entropy of a sphere under the relevant perturbation \cite{Ryu:2006bv,Ryu:2006ef}.
We consider the system holographically described by the $(d+1)$-dimensional Einstein gravity coupled to a free massive scalar field.
Depending on the value of the conformal dimensions, there are two different ways of quantizing the scalar field \cite{Klebanov:1999tb}, one of which was missing in the previous 
consideration \cite{Klebanov:2012va}.
We investigate both cases and find a new behavior of the entanglement entropy for small conformal dimensions.
Indeed, our gravity analysis reveals not only when the stationarity is lost, but also when the perturbative calculation of the entropy breaks down.

Before closing the introduction, we summarize the main results of this letter.\begin{itemize}
\item
For $\frac{d}{2} < \D < d$, the entanglement entropy of a sphere is stationary against the relevant deformation at the UV fixed point. 
Perturbative calculation works with respect to the coupling constant.
\item
For $\frac{d}{3} < \D \le \frac{d}{2}$, the entanglement entropy is still stationary at the UV fixed point, but the perturbative expansion fails.
\item
For $\frac{d}{2}-1 < \D \le \frac{d}{3}$, the entanglement entropy is neither stationary nor perturbative at the UV fixed point.
\end{itemize}
We speculate that these statements hold for any relevant perturbation in any  quantum field theory invariant under the Poincar{\'e} symmetry.

\section{Numerical calculation for free massive scalar in three-dimensions}\label{ss:NumericEE}

A free massive scalar theory whose action is given by
\begin{align}
I = - \frac{1}{2}\int d^3 x \left[ (\partial_\mu \phi)^2 + m^2 \phi^2 \right] \ ,
\end{align}
can be regarded as a relevant deformation of a free massless scalar theory by the operator and the coupling
\begin{align}\label{ScalarOp}
\lambda = m^2 \ , \qquad \CO = \phi^2 \ ,
\end{align}
where the conformal dimension of $\CO$ is $\D = 1$.
Although the analytic computation of the entanglement entropy of the theory has not been known, the numerical studies are conducted in \cite{Klebanov:2012va} where the renormalized entanglement entropy $\CF(R)$ defined by \eqref{REE}
is shown to be non-stationary at the UV fixed point $m^2=0$.\footnote{See also \cite{Huerta:2011qi,Klebanov:2012yf,Safdi:2012sn,Lee:2014xwa} for the studies of the large mass limit.} 

Here we study the behavior of $\CF(R)$ more carefully following the method in \cite{Klebanov:2012va} that is reviewed in appendix \ref{ss:app}.
Figure \ref{fig:massivescalar} shows the numerical plot of the REE of the free massive scalar field with respect to the dimensional coupling $(mR)^2$.
It approaches to $0.0638$ as $m$ goes to zero that coincides with the exact result $\CF_\text{UV} = \frac{1}{16}\left(2\log 2 - 3\frac{\zeta (3)}{\pi^2}\right)\approx 0.0638$ obtained by \cite{Klebanov:2011gs}, while it monotonically decreases to zero as $m$ becomes large in accord with the $F$-theorem \cite{Jafferis:2011zi,Klebanov:2011gs,Casini:2012ei}.
In the right panel we plot the UV region of the left panel.
We find the line tangent to $\CF$ at $m=0$ reads
\begin{align}\label{REELinear}
\CF(mR)= \CF_\text{UV} - 0.133 (mR)^2 \ .
\end{align}
Thus the (renormalized) entanglement entropy of the free massive scalar is not stationary at the UV fixed point.
Furthermore, 
this implies that the perturbative expansion with respect to $\lambda = m^2$ fails.
In the next section, we will study the relevant perturbation of entanglement entropy by using the AdS/CFT correspondence with the bulk massive scalar field dual to the relevant operator.
We will see the same linearity as \eqref{REELinear} is obtained for a general operator with $\D=1$.

\begin{figure}[htbp]
\centering
\includegraphics[width=7.5cm]{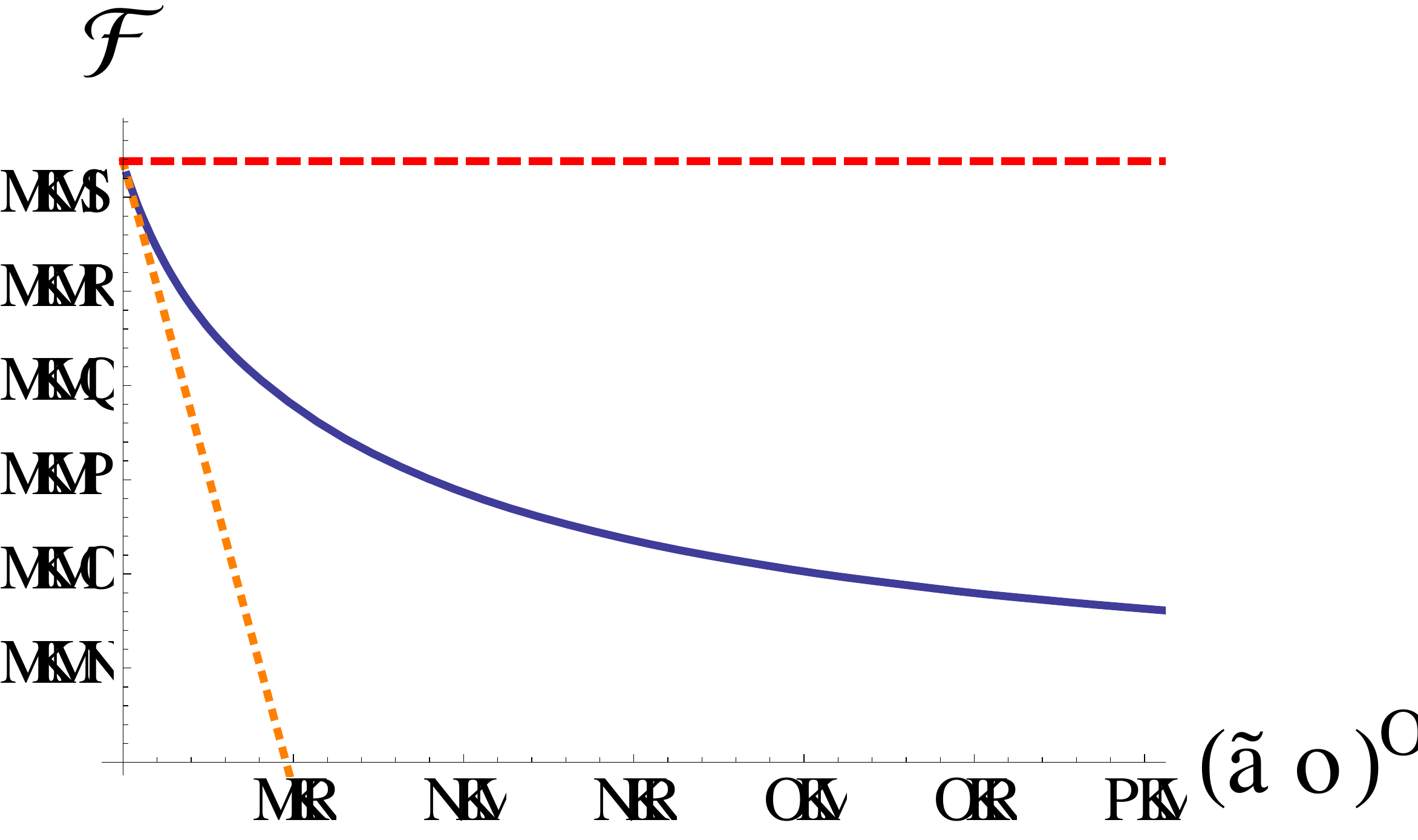}\hspace{0.5cm}
\includegraphics[width=7.5cm]{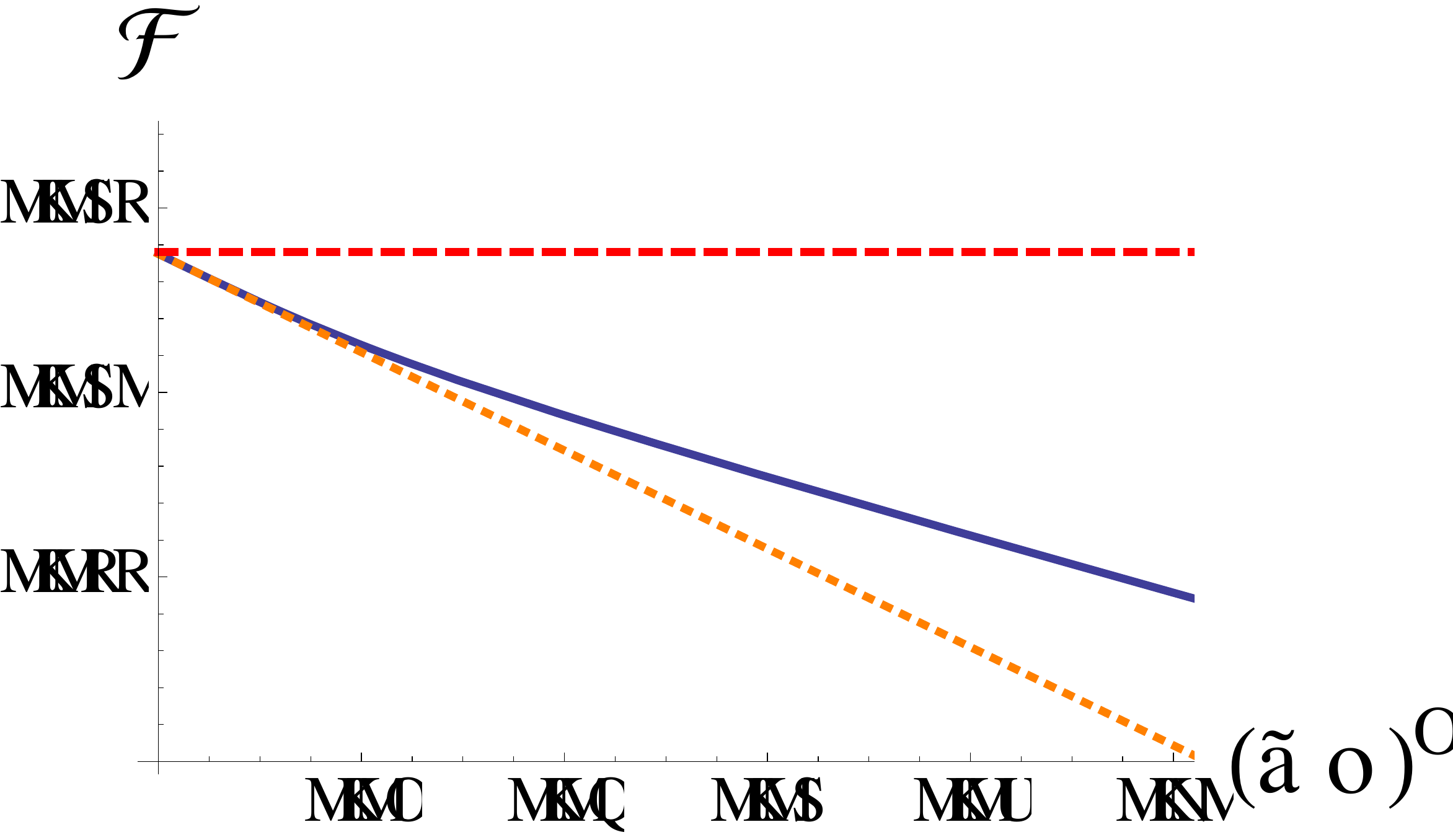}
\caption{The renormalized entanglement entropy of a free massive scalar field of mass $m$ is shown by the blue solid curve. The red dashed line is the value at the UV fixed point, $\CF_\text{UV} = \frac{1}{16}\left(2\log 2 - 3\frac{\zeta (3)}{\pi^2}\right)\approx 0.0638$. The orange dotted line is the tangent curve at the UV fixed point given by $\CF(mR)= \CF_\text{UV} -0.133 (mR)^2 $.}
\label{fig:massivescalar}
\end{figure}

\section{Holographic view of the relevant perturbation}\label{ss:Holographic}

We will examine the holographic entanglement entropy of a sphere of radius $R$ below.
We choose the metric of an asymptotically AdS$_{d+1}$ space as
\begin{align}\label{AAdS}
ds^2 = \frac{L^2}{z^2} \left( -dt^2 + dr^2 + r^2 d\Omega_{d-2}^2 + \frac{dz^2}{f(z)}\right) \ ,
\end{align}
where $f(z)$ approaches to one as $z$ goes to zero.
The relevant perturbation of CFT can be holographically described by a free massive scalar field coupled to the space-time
\begin{align}
I = \frac{1}{16\pi G_N}\int d^{d+1}x \, \sqrt{g} \left[ R - \frac{d(d-1)}{L^2}\right]- \frac{1}{2} \int d^{d+1} x \, \sqrt{g} \left[ (\partial\Phi)^2 + M^2 \Phi^2 \right] \ .
\end{align}
The AdS/CFT correspondence \cite{Maldacena:1997re,Gubser:1998bc,Witten:1998qj} relates  the mass $M$ of the scalar field and the conformal dimension $\D$ of the boundary operator $\CO$:
\begin{align}\label{MassDelta}
\D_\pm = \frac{d}{2} \pm \sqrt{(ML)^2 + \frac{d^2}{4}} \ .
\end{align}
Both $\D_\pm$ are possible for $- \frac{d^2}{4} \le (ML)^2 \le - \frac{d^2}{4} +1$ while only $\D_+$ is allowed for $- \frac{d^2}{4} + 1 < (ML)^2$ \cite{Klebanov:1999tb}.
In the asymptotically AdS$_{d+1}$ space, the scalar field approaches to the boundary $z=0$ as
\begin{align}\label{PhiExpansion}
\Phi (z, \vec x) \to z^{\D_+} [A(\vec x) + \cdots] + z^{\D_-} [B(\vec x) + \cdots ]\ ,
\end{align}
where we can choose $\D$ to be either $\D_+$ or $\D_-$ as long as the mass of the scalar field is in the range mentioned above.
When $\D=d/2$, $\D_+$ and $\D_-$ degenerate and we need to replace $z^{\D_-}$ with $z^{\D_-}\log z$ in \eqref{PhiExpansion}.

Depending on which conformal dimensions we use, there are two ways of quantizations of the bulk scalar field \cite{Klebanov:1999tb}. 
For $\D = \D_+$ (the standard quantization), we regard $A(\vec x)$ and $B(\vec x)$ as the vacuum expectation value of the operator $\langle\CO\rangle$ and the source coupling $\lambda$, respectively.
On the other hand, for $\D = \D_-$ (the alternative quantization), we flip the roles of  $A(\vec x)$ and $B(\vec x)$, i.e. $A(\vec x)$ is identified with $\lambda$ while $B(\vec x)$ with $\langle \CO\rangle$.
Note that we have to use the alternative quantization if $\frac{d}{2}-1 \le \D < \frac{d}{2}$.

Since we are interested in varying the coupling $\lambda$, we only consider the source deformation.
By solving the equations of motion, we can take into account the back reaction of the scalar field to the metric. 
For small $z$, the function $f(z)$ behaves as
\begin{align}
f(z) = 1 + \left\{
	\begin{array}{ll}
	(\mu z)^{2\alpha} + \cdots \ , &\D \neq d/2 \ , \\
	(\mu z)^{d}(\log \mu z)^2 + \cdots \ , & \D = d/2 \ ,
	\end{array}\right.
\end{align}
where $\mu$ is a mass scale determined by the coupling $\lambda$ of the relevant operator.\footnote{Since $\lambda$ is the only dimensionful parameter, $\mu\sim \lambda^{1/(d-\D)}$ from the dimensional analysis.} 
For a source deformation, $\alpha = d-\D_+$ in the standard quantization, and $\alpha = \D_-$ in the alternative quantization.

Now we consider the variation of the entanglement entropy of a sphere of radius $R$ by the relevant perturbation, which is holographically studied in \cite{Liu:2012eea}.
The holographic entanglement entropy of a given entangling surface dividing space into two pieces can be obtained by the area of the minimal surface in the asymptotically AdS$_{d+1}$ space \eqref{AAdS} whose boundary coincides with the entangling surface \cite{Ryu:2006bv,Ryu:2006ef}.
In our case, it is given by minimizing the following functional:
\begin{align}
S = K\int_0^{z_m}dz\, \frac{r(z)^{d-2}}{z^{d-1}}\sqrt{r'(z)^2 + \frac{1}{f(z)}} \ ,\qquad K = \frac{L^{d-1}}{4G_N} \text{Vol}(S^{d-2}) \ , 
\end{align}
where $\text{Vol}(S^{d-2})$ is the volume of the $(d-2)$-dimensional unit sphere, and the boundary conditions are imposed at $z=0$ by $r(0)=R$ and at the tip of the minimal surface $z=z_m$ by $r(z_m)=0,~r'(z_m)=\infty$.\footnote{This is the disk type solution, while there exists the cylinder type solution which extends to $z=\infty$ \cite{Liu:2012eea}. We only consider the former because we are interested in the UV fixed point where only the former solution exists.}
When $f(z)=1$, the minimal surface is a half $(d-1)$-dimensional sphere of radius $R$: $r_0(z) = \sqrt{R^2 - z^2}$.
The variation of the metric $f=1+\delta f$ induces a solution $r = r_0 + \delta r$ with small $\delta r$.
Solving the equation of motion for $\delta r$ with $\delta f = (\mu z)^{2\alpha}$ 
the variation of the entanglement entropy becomes \cite{Liu:2012eea}
\begin{align}\label{Svariation}
\delta S &=\left\{ \begin{array}{ll}
- \frac{\Gamma(\frac{d+1}{2})\Gamma(\frac{2-d+2\alpha}{2})}{4\Gamma(\frac{3}{2}+\alpha)}K  (\mu R)^{2\alpha} 
+ \frac{K (\mu R)^{d-2}}{2(2-d+2\alpha)}(\mu \epsilon)^{2-d+2\alpha} + O(\epsilon^{4-d+2\alpha}) \ ,  & \D\neq d/2 \ ,\\
-\frac{K}{2
   (d+1)}(\mu  R)^d  \log^2
   (\mu R) + O(\epsilon^2 \log^2 (\mu \epsilon))  \ ,  & \D= d/2 \ ,
\end{array}\right.
\end{align}
where $\epsilon$ is the UV cutoff introduced at $z=\epsilon$ and the expansion is given in terms of the dimensionless parameter $\mu R$.
In the standard quantization, the second term for $\D\neq d/2$ can be divergent that needs to be renormalized as well as the UV divergent terms of the unperturbed entropy, while it is finite in the alternative quantization because of $d-2<2\alpha < d$.

We introduce a dimensionless coupling by $t\equiv \lambda R^{d-\D}$ to study how the entanglement entropy varies near the UV fixed point ($t=0$).
It follows from \eqref{Svariation} that the first derivative of the entropy with respect to $t$ is
\begin{align}\label{Sder}
\frac{dS}{dt} =
\left\{\begin{array}{ll}
 - \#\, t^{\frac{2\alpha}{d-\D}-1} + \cdots\ , &\qquad \D\neq d/2 \ ,\\
 - \#\, t \log^2 t +\cdots  \ ,  &\qquad \D= d/2 \ ,
\end{array}\right.
\end{align}
where $\#$ are positive constants we are not interested in.
In the standard quantization ($\D \ge d/2$), the exponent of $t$ is always one and $dS/dt=0$ at $t=0$.
On the other hand, $dS/dt$ can be non-zero in the alternative quantization ($\frac{d}{2}-1 < \D < d/2$) when the conformal dimension of the relevant operator is in the following range:
\begin{align}\label{NSrange}
\frac{d}{2}-1 < \D \le \frac{d}{3} \ .
\end{align}
It follows that such a relevant operator exists in $d\le 6$ dimensions. 

In addition to the stationarity, we remark the condition for the perturbative expansion being well-defined.
The expansion near the UV fixed point \eqref{Svariation} or \eqref{Sder} indicates an interesting fact such that 
the exponent of the dimensionless coupling $t$ of the leading term of the entropy is always two for the standard quantization with $\D > d/2$, but it can be non-integers for the alternative quantization with $\D \le d/2$.
It implies the failure of the perturbation theory, and we are led to the statements in the introduction.

Let us compare our findings with existing examples in field theories.
The numerical observation in section \ref{ss:NumericEE} yields $\delta S \sim t$ with $t = (mR)^2$ for the relevant operator \eqref{ScalarOp} of $\D =1$.
The gravity result \eqref{Sder} explains the linearity since it saturates the upper bound of \eqref{NSrange} when $d=3$.

Another example of interest is a free massive fermion in two-dimensions \cite{Casini:2005rm,Herzog:2013py} whose entanglement entropy takes the form of $\delta S \sim t^2 \log^2 t$ with $t= m R$ near the UV fixed point.
Since the mass deformation has $\D = 1 = d/2$, it agrees with \eqref{Sder} up to a numerical factor.
The non-analyticity of the entropy emanates from the IR divergence \cite{Herzog:2013py}, and leads to the break down of the perturbative calculation.

 \vspace{1.3cm}
 \centerline{\bf Acknowledgements}
I am grateful to T.\,Bargheer, H.\,Casini, M.\,Huerta, I.\,Klebanov, J.\,Maldacena, Y.\,Nakaguchi, S.\,Pufu and M.\,Rangamani for valuable discussions, and especially to K.\,Yonekura for helpful discussions and comments on the manuscript of this letter.
This work was supported by a JSPS postdoctoral fellowship for research abroad.

\appendix

\section{The numerics of entanglement entropy }\label{ss:app}
In this appendix, we review the numerical computation of entanglement entropy of a disk of radius $R$ for a free massive scalar field in three-dimensions used in section \ref{ss:NumericEE} following \cite{Srednicki:1993im,Huerta:2011qi,Klebanov:2012va,Lee:2014xwa}. 

A free scalar field can be Fourier decomposed into modes with angular momentum $n$ ($n=0,1,2,\cdots$) in the polar coordinates \eqref{flat}.
To put the theory on a lattice, we discretize the radial coordinate $r$ to $N$ points labeled by $i=1,\cdots, N$.
Then the Hamiltonian on the lattice becomes
\begin{align}
H = \frac{1}{2}\sum_{n=0}^\infty \left[ \sum_{i=1}^N \pi_{n,i}^2 + \sum_{i,j=1}^N \phi_{n,i} K_n^{i,j}\phi_{n,j} \right] \ ,
\end{align}
where $\pi_{n,i}$ are the $n$-th momentum on the $i$-th site on the lattice conjugate to the discretized scalar field $\phi_{n,i}$.
For a free massive scalar field of mass $m^2$, the matrices $K_n^{i,j}$ are given by
\begin{align}
K_n^{1,1} = \frac{3}{2} + n^2 + m^2 \ , \qquad K_n^{i,i} = 2 +\frac{n^2}{i^2} + m^2 \ , \qquad K_n^{i,i+1} = K_n^{i+1,i} = - \frac{i+1/2}{\sqrt{i(i+1)}} \ .
\end{align}
These matrices are related to the two-point functions of the scalar fields and the conjugate momenta as $(X_n)_{ij} = \langle \phi_{n,i}\phi_{n,j} \rangle = \frac{1}{2}(K^{-1/2}_n)_{ij}$ and $(P_n)_{ij} = \langle \pi_{n,i}\pi_{n,j} \rangle = \frac{1}{2}(K^{1/2}_n)_{ij}$, respectively.
Let the radius of the disk be a half-integer in units of the lattice spacing, $R=r + 1/2$ with integer $r$. 
To calculate the entanglement entropy inside the disk, we take $r\times r$ submatrices denoted by $(X_n^r)_{ij}$ and $(P_n^r)_{ij}$ of the matrices $X_n,P_n$ with the restricted ranges $1\le i,j\le r$.
Then the entropy is obtained by
\begin{align}\label{NEE}
S(R) = S_0 + 2\sum_{n=1}^\infty S_n \ , 
\end{align}
where 
\begin{align}
S_n = \tr \left[ \left( C_n + 1/2\right) \log\left( C_n + 1/2\right) - \left( C_n - 1/2\right) \log\left( C_n - 1/2\right) \right] \ ,
\end{align}
with $C_n\equiv \sqrt{X_n^r P_n^r}$. It is positive because the eigenvalues of $C_n$ are equal or bigger than $1/2$.

To evaluate \eqref{NEE} numerically, we take a lattice of $N=200$ points and change the radius between $30 < r < 50$.
The mass parameter is varied as $m=0.002 \cdot a$ with $a=1,\cdots, 20$.
To treat the finite lattice effects and the summation over an infinite number of the angular modes, we carry out the calculation in the following ways:
\begin{itemize}
\item
For $6\le n \le 2000$, the entropies are summed over $n$ with the prescription mentioned above.
\item
For $0 \le n \le 5$, we repeat the calculations by changing the sizes of the lattice as $N=200 + 10 \cdot b$ with $b=0,1,\cdots, 50$.
Then we fit the results to \cite{Huerta:2011qi}
\begin{align}
S_n = s_n + \frac{a_2}{N^2} + \frac{b_2 \log N}{N^2}+ \frac{a_4}{N^4} + \frac{b_4 \log N}{N^4} + \frac{a_6}{N^6} + \frac{b_6 \log N}{N^6} \ ,
\end{align}
and read off the term $s_n$ as the values of $S_n$ in the $N\to \infty$ limit.
\item
For $2001\le n$, we use the large $n$ expansions of $S_n$ derived in \cite{Klebanov:2012va}
\begin{align}
S_n = c_n (1-\log c_n) + \frac{(m^2 +2)(2r+1)^2}{2n^2}c_n \log c_n + O(\log n/n^8) \ ,
\end{align}
with $c_n= r^2 (r+1)^2/(16n^4)$.
\end{itemize}


\bibliographystyle{JHEP}
\bibliography{PerturbativeEE}

\providecommand{\href}[2]{#2}\begingroup\raggedright\begin{thebibliography}{10}

\bibitem{Zamolodchikov:1986gt}
A.~B. Zamolodchikov, {\it {Irreversibility of the Flux of the Renormalization
  Group in a 2D Field Theory}},  {\em JETP Lett.} {\bf 43} (1986) 730--732.

\bibitem{Casini:2004bw}
H.~Casini and M.~Huerta, {\it {A Finite Entanglement Entropy and the
  C-Theorem}},  {\em Phys.Lett.} {\bf B600} (2004) 142--150,
  [\href{http://xxx.lanl.gov/abs/hep-th/0405111}{{\tt hep-th/0405111}}].

\bibitem{Jafferis:2011zi}
D.~L. Jafferis, I.~R. Klebanov, S.~S. Pufu, and B.~R. Safdi, {\it {Towards the
  F-Theorem: ${\mathcal{N}}\!=2$ Field Theories on the Three-Sphere}},  {\em
  JHEP} {\bf 1106} (2011) 102, [\href{http://xxx.lanl.gov/abs/1103.1181}{{\tt
  1103.1181}}].

\bibitem{Klebanov:2011gs}
I.~R. Klebanov, S.~S. Pufu, and B.~R. Safdi, {\it {F-Theorem without
  Supersymmetry}},  \href{http://xxx.lanl.gov/abs/1105.4598}{{\tt 1105.4598}}.

\bibitem{Casini:2012ei}
H.~Casini and M.~Huerta, {\it {On the RG Running of the Entanglement Entropy of
  a Circle}},  {\em Phys.Rev.} {\bf D85} (2012) 125016,
  [\href{http://xxx.lanl.gov/abs/1202.5650}{{\tt 1202.5650}}].

\bibitem{Liu:2012eea}
H.~Liu and M.~Mezei, {\it {A Refinement of Entanglement Entropy and the Number
  of Degrees of Freedom}},  {\em JHEP} {\bf 1304} (2013) 162,
  [\href{http://xxx.lanl.gov/abs/1202.2070}{{\tt 1202.2070}}].

\bibitem{Lieb:1973cp}
E.~Lieb and M.~Ruskai, {\it {Proof of the Strong Subadditivity of
  Quantum-Mechanical Entropy}},  {\em J.Math.Phys.} {\bf 14} (1973) 1938--1941.

\bibitem{Myers:2010tj}
R.~C. Myers and A.~Sinha, {\it {Holographic C-Theorems in Arbitrary
  Dimensions}},  \href{http://xxx.lanl.gov/abs/1011.5819}{{\tt 1011.5819}}.

\bibitem{Jafferis:2012iv}
D.~L. Jafferis and S.~S. Pufu, {\it {Exact Results for Five-Dimensional
  Superconformal Field Theories with Gravity Duals}},
  \href{http://xxx.lanl.gov/abs/1207.4359}{{\tt 1207.4359}}.

\bibitem{Fei:2014yja}
L.~Fei, S.~Giombi, and I.~R. Klebanov, {\it {Critical $O(N)$ Models in
  6-$\epsilon$ Dimensions}},  \href{http://xxx.lanl.gov/abs/1404.1094}{{\tt
  1404.1094}}.

\bibitem{Klebanov:2012va}
I.~R. Klebanov, T.~Nishioka, S.~S. Pufu, and B.~R. Safdi, {\it {Is Renormalized
  Entanglement Entropy Stationary at RG Fixed Points?}},  {\em JHEP} {\bf 1210}
  (2012) 058, [\href{http://xxx.lanl.gov/abs/1207.3360}{{\tt 1207.3360}}].

\bibitem{Casini:2005zv}
H.~Casini and M.~Huerta, {\it {Entanglement and Alpha Entropies for a Massive
  Scalar Field in Two Dimensions}},  {\em J. Stat. Mech.} {\bf 0512} (2005)
  P12012, [\href{http://xxx.lanl.gov/abs/cond-mat/0511014}{{\tt
  cond-mat/0511014}}].

\bibitem{Casini:2009sr}
H.~Casini and M.~Huerta, {\it {Entanglement Entropy in Free Quantum Field
  Theory}},  {\em J. Phys.} {\bf A42} (2009) 504007,
  [\href{http://xxx.lanl.gov/abs/0905.2562}{{\tt 0905.2562}}].

\bibitem{Rosenhaus:2014woa}
V.~Rosenhaus and M.~Smolkin, {\it {Entanglement Entropy: a Perturbative
  Calculation}},  \href{http://xxx.lanl.gov/abs/1403.3733}{{\tt 1403.3733}}.

\bibitem{Kim:2014yca}
K.~K. Kim, O.-K. Kwon, C.~Park, and H.~Shin, {\it {Non-Stationary Entanglement
  Entropy Flow in Mass-Deformed Abjm Theory}},
  \href{http://xxx.lanl.gov/abs/1404.1044}{{\tt 1404.1044}}.

\bibitem{Lin:2004nb}
H.~Lin, O.~Lunin, and J.~M. Maldacena, {\it {Bubbling AdS Space and 1/2 BPS
  Geometries}},  {\em JHEP} {\bf 0410} (2004) 025,
  [\href{http://xxx.lanl.gov/abs/hep-th/0409174}{{\tt hep-th/0409174}}].

\bibitem{Ryu:2006bv}
S.~Ryu and T.~Takayanagi, {\it {Holographic Derivation of Entanglement Entropy
  from AdS/CFT}},  {\em Phys. Rev. Lett.} {\bf 96} (2006) 181602,
  [\href{http://xxx.lanl.gov/abs/hep-th/0603001}{{\tt hep-th/0603001}}].

\bibitem{Ryu:2006ef}
S.~Ryu and T.~Takayanagi, {\it {Aspects of Holographic Entanglement Entropy}},
  {\em JHEP} {\bf 08} (2006) 045,
  [\href{http://xxx.lanl.gov/abs/hep-th/0605073}{{\tt hep-th/0605073}}].

\bibitem{Klebanov:1999tb}
I.~R. Klebanov and E.~Witten, {\it {AdS / CFT Correspondence and Symmetry
  Breaking}},  {\em Nucl.Phys.} {\bf B556} (1999) 89--114,
  [\href{http://xxx.lanl.gov/abs/hep-th/9905104}{{\tt hep-th/9905104}}].

\bibitem{Huerta:2011qi}
M.~Huerta, {\it {Numerical Determination of the Entanglement Entropy for Free
  Fields in the Cylinder}},  \href{http://xxx.lanl.gov/abs/1112.1277}{{\tt
  1112.1277}}.

\bibitem{Klebanov:2012yf}
I.~R. Klebanov, T.~Nishioka, S.~S. Pufu, and B.~R. Safdi, {\it {On Shape
  Dependence and RG Flow of Entanglement Entropy}},  {\em JHEP} {\bf 1207}
  (2012) 001, [\href{http://xxx.lanl.gov/abs/1204.4160}{{\tt 1204.4160}}].

\bibitem{Safdi:2012sn}
B.~R. Safdi, {\it {Exact and Numerical Results on Entanglement Entropy in
  (5+1)-Dimensional CFT}},  {\em JHEP} {\bf 1212} (2012) 005,
  [\href{http://xxx.lanl.gov/abs/1206.5025}{{\tt 1206.5025}}].

\bibitem{Lee:2014xwa}
J.~Lee, L.~McGough, and B.~R. Safdi, {\it {Renyi Entropy and Geometry}},
  \href{http://xxx.lanl.gov/abs/1403.1580}{{\tt 1403.1580}}.

\bibitem{Maldacena:1997re}
J.~M. Maldacena, {\it {The Large $N$ Limit of Superconformal Field Theories and
  Supergravity}},  {\em Adv.Theor.Math.Phys.} {\bf 2} (1998) 231--252,
  [\href{http://xxx.lanl.gov/abs/hep-th/9711200}{{\tt hep-th/9711200}}].

\bibitem{Gubser:1998bc}
S.~S. Gubser, I.~R. Klebanov, and A.~M. Polyakov, {\it {Gauge Theory
  Correlators from Non-Critical String Theory}},  {\em Phys. Lett.} {\bf B428}
  (1998) 105--114, [\href{http://xxx.lanl.gov/abs/hep-th/9802109}{{\tt
  hep-th/9802109}}].

\bibitem{Witten:1998qj}
E.~Witten, {\it {Anti-de~Sitter Space and Holography}},  {\em Adv. Theor. Math.
  Phys.} {\bf 2} (1998) 253--291,
  [\href{http://xxx.lanl.gov/abs/hep-th/9802150}{{\tt hep-th/9802150}}].

\bibitem{Casini:2005rm}
H.~Casini, C.~D. Fosco, and M.~Huerta, {\it {Entanglement and Alpha Entropies
  for a Massive Dirac Field in Two Dimensions}},  {\em J. Stat. Mech.} {\bf
  0507} (2005) P07007, [\href{http://xxx.lanl.gov/abs/cond-mat/0505563}{{\tt
  cond-mat/0505563}}].

\bibitem{Herzog:2013py}
C.~P. Herzog and T.~Nishioka, {\it {Entanglement Entropy of a Massive Fermion
  on a Torus}},  {\em JHEP} {\bf 1303} (2013) 077,
  [\href{http://xxx.lanl.gov/abs/1301.0336}{{\tt 1301.0336}}].

\bibitem{Srednicki:1993im}
M.~Srednicki, {\it {Entropy and Area}},  {\em Phys. Rev. Lett.} {\bf 71} (1993)
  666--669, [\href{http://xxx.lanl.gov/abs/hep-th/9303048}{{\tt
  hep-th/9303048}}].

\end{thebibliography}\endgroup

\end{document}